\documentclass{article}

\usepackage[preprint]{neurips_2026}

\usepackage[utf8]{inputenc} 
\usepackage[T1]{fontenc}    
\usepackage{hyperref}       
\usepackage{url}            
\usepackage{booktabs}       
\usepackage{amsfonts}       
\usepackage{nicefrac}       
\usepackage{microtype}      
\usepackage{xcolor}         

\usepackage{longtable, booktabs} 
\newcommand{\parnobf}[1]{\par \noindent {\bf #1}}

\newcommand{\ai}{\textit{Clinician + AI}}
\newcommand{\alone}{\textit{Clinician Alone}}
\newcommand{\spspe}{\textit{Clinician + Selective Prediction}}

\usepackage{xcolor}
\usepackage{soul}
\usepackage{graphicx}

\usepackage{subcaption}

\usepackage{xcolor}
\usepackage{booktabs}
\usepackage{multirow} 
\definecolor{custompurple}{RGB}{75, 0, 146} 

\title{Selective Prediction Reduces the Negative Effects of Automation Bias Overall but Increases False Negatives}

\author{%
  \textbf{Sarah Jabbour\textsuperscript{1}, David Fouhey\textsuperscript{2},
  Nikola Banovic\textsuperscript{1}, Stephanie D. Shepard\textsuperscript{1},} \\
  \textbf{Ella Kazerooni\textsuperscript{1},
  Michael W. Sjoding\textsuperscript{1,$*$},
  Jenna Wiens\textsuperscript{1,$*$}} \\
  \vspace{0.5em} \\
  \begin{minipage}{0.9\textwidth}
    \centering
    \textsuperscript{1}University of Michigan,
    \textsuperscript{2}New York University \\
    \vspace{0.3em}
    \texttt{\{sjabbour, wiensj\}@umich.edu} \\
    \textsuperscript{$*$}Equal contribution.
  \end{minipage}
}

\begin{document}

\maketitle

\begin{abstract}
         High-performing models can sometimes produce inaccurate predictions and hurt human decisions when deployed. \textit{Selective prediction}, in which potentially harmful model predictions are hidden from users, has been proposed as a solution. Typical evaluations of this approach rely on simulations that assume that when AI abstains (and informs the user so) humans make decisions as they would without AI involvement in their workflow, but this assumption has not been validated in multilabel tasks. To test this assumption, we study the effects of selective prediction on human decisions in a clinical context. We conducted a vignette-based study of 259 clinicians tasked with treating hospitalized patients. We compared their baseline performance without \textit{any} AI involvement in their workflow to their AI-assisted performance \textit{with} and \textit{without} selective prediction. Our findings indicate that selective prediction partly mitigates the negative effects of inaccurate AI in terms of decision accuracy. However, it alters the type of errors clinicians make: when informed that the AI abstains, clinicians' false negative rate is significantly elevated (i.e., they undertreat patients), particularly those who have not previously used AI in their clinical practice. Our findings underscore the need to evaluate AI systems in real-world workflows. 
\end{abstract}

\section{Introduction}

AI models have been developed to assist human decision-making in a variety of domains, such as child welfare~\citep{de2020case}, wildlife conservation~\citep{beery2019efficient}, and healthcare~\citep{nori2023capabilities,gulshan2016development,adams2022prospective,mckinney2020international}. Yet, as AI-assisted decision-making becomes more prevalent, the risk of harm from incorrect AI predictions also grows. For example, automation bias, in which humans overrely on incorrect AI predictions, can propagate harmful model errors. This is especially worrisome in high-stakes settings, such as healthcare, where it can result in incorrect diagnoses ~\citep{jabbour2023measuring,gaube2021ai}. As a result, improving AI-assisted decision-making requires not only more accurate models, but careful design of how and when AI predictions are presented to humans.

\begin{figure*}[!t]
    \centering
    \includegraphics[width = \textwidth]{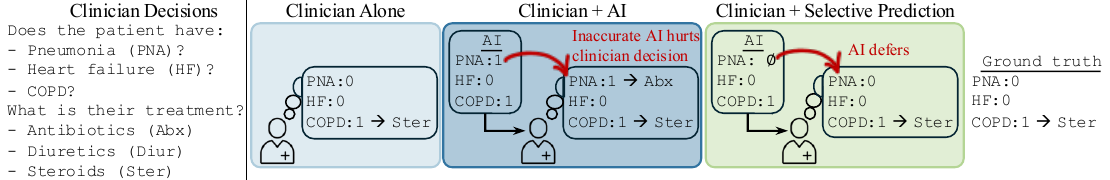}
   \caption{\textbf{Three types of clinician decision-making workflows with varying levels of AI input.} For each patient, clinicians are tasked with determining their diagnoses and treatments. In the \alone{} setting, clinicians receive no input from AI. In the \ai{} setting, an AI provides predictions for all conditions, which can harm clinical decisions when the model is wrong. In the \spspe{} setting, the AI defers on a subset of predictions, showing only potentially helpful predictions while withholding those that may be harmful.}
    \label{fig:selective_prediction}
\end{figure*}

One increasingly popular design strategy is \textit{selective prediction}~\citep{el2010foundations}, also known as AI deferral~\citep{mozannar2023should}. In selective prediction, an AI system withholds its prediction on cases where showing them may harm human decision-making and instead asks the human to decide on their own. In contrast to the standard \ai{} workflow, where the AI provides input for every decision (Figure~\ref{fig:selective_prediction}), selective prediction assumes the existence of some mechanism that identifies potentially harmful predictions (e.g., incorrect or uncertain predictions). For these cases, the system abstains from showing the prediction and asks the human to make a decision on their own (Figure~\ref{fig:selective_prediction} \spspe). This design is intended to reduce overreliance by preventing exposure to harmful AI outputs. A growing body of work has proposed algorithms for learning when to defer to the human~\citep{lykouris2024learning,mozannar2023should,li2024joint,shah2022selective,chen2023aspest,nam2022selective,strong2025trustworthy}. 

Despite such algorithmic progress, the effects of selective prediction on human decision-making remain underexplored. For the most part, selective prediction has been evaluated in simulated scenarios~\citep{nam2022selective,farzaneh2023collaborative,wang2023artificial}. Here, joint human-AI performance is calculated using retrospective data. Independent decisions are first collected from both humans and AI on all cases, and then various human-AI teaming strategies are simulated to calculate humans' AI-assisted performance. Crucially, \textit{this relies on the behavioral assumption that when the AI abstains (and informs the user so), the human will respond as if no AI were involved in their workflow at all}. However, a key distinction exists between workflows with no AI and those in which the human is told that the AI has withheld its output. Prior work by ~\cite{bondi2022role} empirically studied AI deferral in human decision-making, but focused on a single-output setting in which AI input is either fully revealed or completely withheld. In contrast, we study a multi-output setting in which the AI can selectively reveal some outputs while abstaining on others. We hypothesize that informing the human that the AI withheld a subset of its predictions can itself alter human decision-making.

In this work, we test this hypothesis in a multilabel clinical decision-making setting as a case study. We ask the question: \textbf{when an AI system selectively withholds predictions and explicitly informs users that it has abstained, do humans behave as they would in a workflow with no AI involvement at all?}
To answer this, we conducted a study of 259 clinicians tasked with treating the causes of hospitalized patients' acute respiratory failure (shortness of breath). Acute respiratory failure (ARF) is a common condition that is often treated incorrectly and could benefit from AI-augmented clinical decision-making \citep{kempker2020epidemiology,zwaan2012relating,jabbour2022combining}. ARF provides an informative testbed because patients often have \textit{multiple concurrent causes and require multiple treatments}. The multilabel structure allows selective prediction to hide some predictions while revealing others, which mirrors realistic clinical deployment. We recruited clinicians who commonly care for these patients to complete a vignette-based survey, randomizing them to receive AI predictions with or without selective prediction, and compared their AI-assisted performance to a no-AI baseline.
\newline 
\parnobf{Contributions:} We present the first study of selective prediction in a multilabel clinical decision-making task. In this study: 

\begin{enumerate}
    \item We find that selective prediction partly mitigates the negative effects of inaccurate AI on human decisions in terms of treatment accuracy. 
    \item We demonstrate that selective prediction changes human errors: when informed that the AI abstains, clinicians make significantly more false negative errors. This can lead to the undertreatment of patients who require intervention. 
\end{enumerate}

\noindent AI deferral to humans is already being framed as a ``first-class safety mechanism''~\citep{azad2026principles}, where clinicians can maintain their autonomy when the AI is wrong or uncertain~\citep{zwaan2026trust}. However, our results show that such deferral mechanisms are not necessarily neutral. In our study, AI abstention shifted clinicians towards undertreatment by increasing false negative treatment errors. This tradeoff is clinically important: in the context of treating ARF, undertreatment (false negatives) may actually pose more harm than overtreatment (false positives)~\citep{zasowski2020systematic}, suggesting that selective prediction should not be treated as a default safety mechanism. Instead, the effects of selective prediction should be evaluated in each end-use context, with careful attention to which types of errors are most acceptable.

\section{Study Task \& Preliminaries} 

In this study, we design an experiment to test the effects of selective prediction on clinician decision-making using simulated patient vignettes grounded in a clinical decision-making task: treating the causes of patients' acute respiratory failure (ARF). ARF has three common causes: pneumonia, heart failure, and chronic obstructive pulmonary disease (COPD). A patient can have any combination of these diagnoses, including all three or none. Thus, clinicians must make multiple simultaneous treatment decisions for a single patient case, where the treatment depends on the diagnosis of the patient. This naturally induces a multilabel decision-making setting in which an AI system generates separate predictions for each possible cause (Figure~\ref{fig:selective_prediction}). 

In an AI-assisted workflow (Figure~\ref{fig:selective_prediction} \ai), the AI’s predictions for all three conditions are shown to the clinician. This risks causing harm if some predictions are inaccurate, and clinicians rely on them. Under selective prediction (Figure~\ref{fig:selective_prediction} \spspe), such potentially harmful predictions were withheld. The clinician is informed that the AI has abstained on some subset and is asked to make a decision on their own. Note that this is explicitly different from the setting where an AI is not part of the clinician's workflow at all (Figure~\ref{fig:selective_prediction} \alone), where the clinician makes a decision on their own for \textit{all} conditions. 

Our goal is to study how selectively withholding AI predictions impacts treatment decisions, independent of the reasons for which the selective prediction mechanism decides to withhold its predictions. To do so, we assume access to a selective prediction mechanism that identifies potentially harmful predictions. We describe this mechanism in Section \ref{experimental_setup}.
\begin{figure*}[!t]
    \centering
    \includegraphics[width=\textwidth]{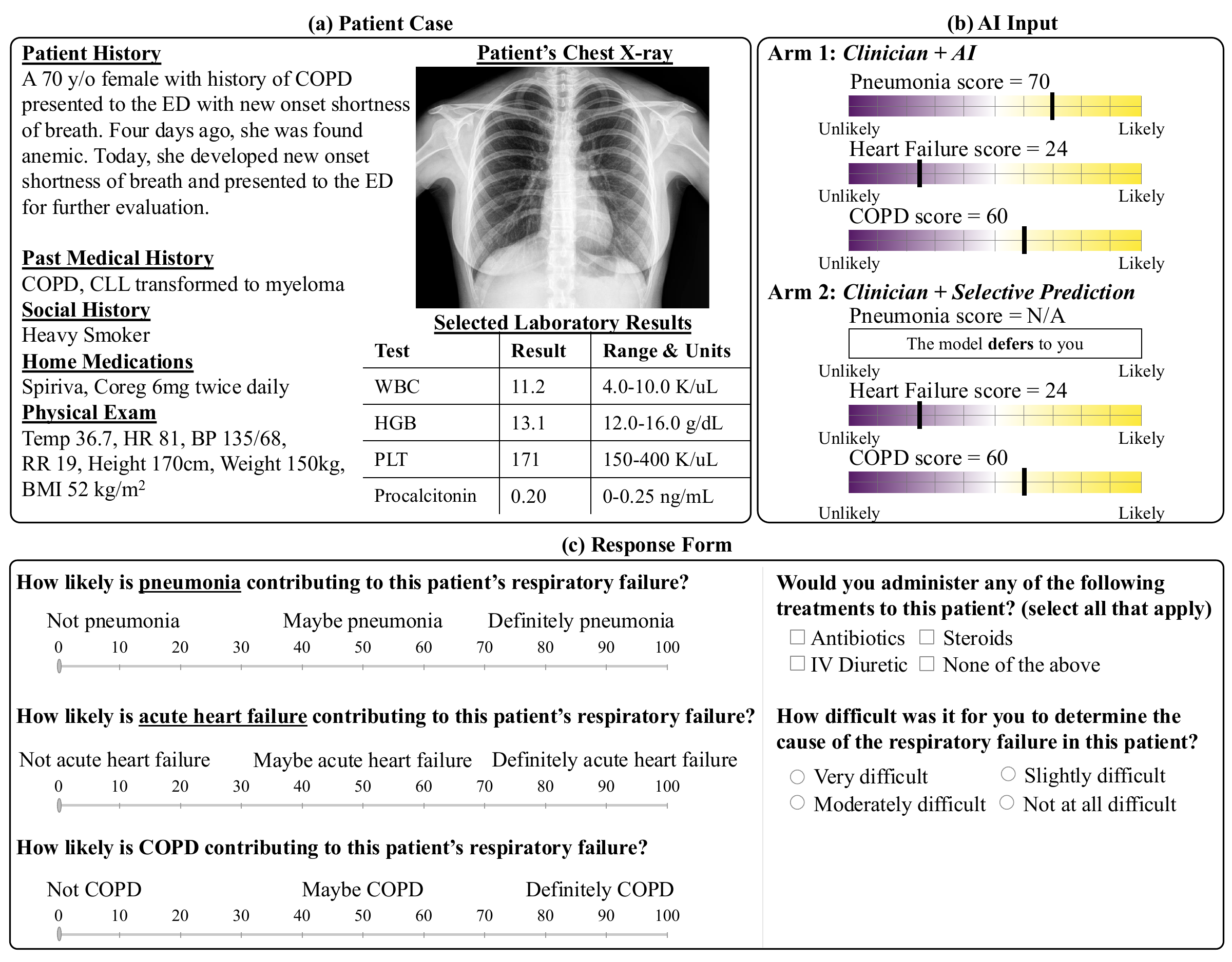}

    \caption{\textbf{Information shown to clinicians during the survey.} (a) Example of how a patient case for an individual hospitalized with ARF was shown to participants, including all information needed to make a diagnosis. Note: this is a fake case shown to preserve patient privacy. (b) How AI model scores were presented to participants in the \ai~ and \spspe~ randomization arms.  (c) Survey response form, where participants diagnosed and treated the patient and rated their perceived difficulty of the case.}
    \label{fig:vignette}
\end{figure*}

\section{Experimental Setup}
\label{experimental_setup}

We conducted a pre-registered study in which we specified the hypotheses, study design, and analysis plan before data collection.\footnote{Pre-registration: \url{https://osf.io/m7avk}} Pre-registration is important because it makes clear which outcomes and analyses were chosen before seeing the data, reducing the risk that results reflect selective reporting rather than a real effect~\citep{chan2004empirical}. We recruited 259 clinicians to test the effects of selective prediction on treatment decisions. The study was designed to isolate behavioral responses to selective prediction by comparing clinician performance under three conditions: no AI in the clinicians' workflow (\alone), all AI predictions given the clinician (\ai), and AI with selective prediction, where a subset of potentially harmful predictions was withheld (\spspe). This study was deemed exempt from IRB review. 

\parnobf{Dataset.}  We used 45 cases of patients hospitalized with ARF at a large academic medical center (see Appendix \ref{patient_cases}). These cases were selected to ensure clinical realism, rather than to train or benchmark the AI model. Ground-truth diagnoses were based on retrospective chart review by multiple physicians. The appropriate treatment for each patient was the standard treatment for each condition: antibiotics for pneumonia, intravenous diuretics for heart failure, and steroids for COPD. 

\parnobf{Study Task.} For each case, clinicians first reviewed the patient's case (Figure \ref{fig:vignette}a). These cases contained clinical details: patient presentation, past medical history, current medications, physical exam findings, laboratory findings, and chest X-ray. Patient data were modified slightly to protect patient privacy without compromising information needed to make a diagnosis (see Appendix \ref{patient_cases}). 

\parnobf{AI Model \& Selective Prediction.}  After reviewing the patient case information, clinicians received AI predictions if applicable  (i.e., \ai{} \& \spspe{} experimental settings)  (Figure \ref{fig:vignette}b). We follow the setup of~\citep{jabbour2022combining}, where the AI model was trained to predict the likelihood that each patient had pneumonia, heart failure, and COPD using their chest X-ray, laboratory results,  vital signs, and demographics. It made 3 predictions for each of the 45 patients  (pneumonia, heart failure, and COPD), assigning a 0-100 score for each condition. The AI's predictions were recalibrated based on disease prevalence so that a threshold of 50 consistently indicated a positive diagnosis across all three outcomes. This was communicated through the colorbar and text ``unlikely" and ``likely" at the far ends. See Appendix \ref{model_details} for further details.

In our experiments, we implement selective prediction using post hoc ground-truth diagnoses. For each patient and each possible diagnosis (pneumonia, heart failure, COPD), we label the AI’s prediction as potentially helpful or potentially harmful based on whether it matches the known ground-truth. These labels determine whether a prediction is shown to the clinician or withheld under selective prediction. To better reflect realistic deployment, where selective prediction mechanisms are unlikely to be perfect, we introduce controlled noise into this process. The mechanism withholds all incorrect predictions. In addition, it withholds a subset of correct predictions so that correct predictions constitute $p$ percent of all withheld predictions. We set $p=0.11$ in our experiments, meaning that approximately 11\% of withheld predictions are correct. We refer to the subset of cases that the mechanism identifies as having potentially harmful predictions as the \textit{inaccurate subset}. Full details on how we identified such predictions and the resulting counts are provided in Appendix~\ref{app:subset_construction}.

\parnobf{Participant Responses.} After reviewing the patient case, clinicians were asked to assess how likely pneumonia, heart failure, and COPD were each independently contributing to the patient’s ARF on a scale from 0 to 100 (the patient can have multiple diagnoses at once) (Figure \ref{fig:vignette}c). After making a diagnosis, clinicians selected what (if any) treatments they would administer (antibiotics, intravenous diuretics, steroids) and reported the perceived difficulty of each patient case on a four-point Likert scale from ``Not at all difficult" to ``Very difficult" (Figure \ref{fig:vignette}c). At the end of the survey, participants completed optional demographic and AI-use related questions (see Appendix \ref{end_of_survey_demographics} for details).


\section{Survey Design \& Randomization} 

The survey was structured to first measure clinicians’ baseline performance without any AI, and then measure how their treatment decisions changed when AI predictions were introduced with or without selective prediction. Each clinician was assigned nine patient cases sampled from the 45 cases (Figure \ref{fig:survey_flow}) (see Appendix \ref{randomization}). The survey presented the nine cases as follows:

\begin{itemize}
    \item \textbf{Cases 1-3: No AI in Workflow} (\alone). The first three cases, randomly chosen from all 45 cases, were presented without AI input. These cases allowed us to measure the clinicians' baseline treatment accuracy without AI input (Figure \ref{fig:survey_flow} `Cases 1-3'). 
    
    \item \textbf{Cases 4–9: AI Input}. The next six cases included AI input. Three were drawn from the \textit{inaccurate subset}, in which the model produced a potentially harmful prediction for at least one condition (see Appendix~\ref{randomization} for details). The remaining three were not from the inaccurate subset, reflecting the fact that clinicians would not encounter selective prediction for \textit{every} patient case in practice. We balanced the case mix to ensure sufficient power to evaluate responses to the \textit{inaccurate subset}, and all six cases were presented in random order to prevent ordering effects. At this stage, participants were randomized into two groups: seeing \textit{all} AI predictions or seeing AI with selective prediction (Figure \ref{fig:survey_flow} `Cases 4-9'). 

    \begin{enumerate}
        \item \textbf{AI without selective prediction} (\ai): Predictions for all conditions were shown, regardless of their accuracy. This allowed us to evaluate the impact of inaccurate predictions on clinical decisions. We hypothesized that clinician accuracy would decrease when seeing inaccurate predictions due to overreliance on inaccurate AI. 

        \item \textbf{AI with selective prediction} (\spspe): The AI model abstained from showing inaccurate predictions, instead presenting the message ``the model \textbf{defers} to you" (Figure \ref{fig:vignette}b). All other predictions were displayed. This allowed us to evaluate the effect of selective prediction in a multilabel setting. We hypothesized that clinician accuracy would not decrease compared to their baseline, as the inaccurate predictions were hidden from them.  
    \end{enumerate}
\end{itemize}

\parnobf{AI Introduction \& Knowledge Check.} After the three \alone{} cases, participants were introduced to the AI model (Figure \ref{fig:survey_flow}). They were told they would see a clinical decision support tool designed to determine the causes of ARF. To avoid bias, they first saw how the tool would look if the task were to identify objects in consumer photographs (Appendix Figure \ref{fig:instructions}). To ensure participants understood selective prediction, those in the \spspe{} randomization arm had to confirm their understanding of selective prediction by selecting a box saying the AI "may or may not believe the patient has the condition, but is choosing not to provide its input (Appendix Figure \ref{fig:knowledge_check}). I should make an assessment on my own." Without this, they were unable to proceed with the survey.

\parnobf{Study Deployment.} The study was deployed on Qualtrics, a web-based survey platform. We recruited clinicians from hospitals across nine US states (see Appendix \ref{survey}).  Participants were clinicians who manage hospitalized patients with ARF and are potential end users of such an AI tool.

\section{Evaluation} 

Our primary goal is to understand how AI, with and without selective prediction, affects human decision-making relative to a workflow with no AI. We evaluate this by comparing clinician performance in \ai{} and \spspe{} to \alone, considering not only overall accuracy but also how selective prediction changes the types of errors clinicians make. Our primary preregistered outcome was diagnostic accuracy. In the survey, each diagnosis mapped directly onto a corresponding treatment decision, and treatments ultimately affect patient outcomes. Thus, in the main text, we present the effects of selective prediction on clinicians' treatment decisions. 

We analyze clinicians’ responses on the \textit{inaccurate subset}, (i.e., conditions for which the selective prediction mechanism identified potentially harmful predictions) across three settings: baseline decisions without AI (\alone), decisions made when clinicians were shown inaccurate AI predictions (\ai), and decisions made when the AI withheld its inaccurate prediction and explicitly abstained (\spspe). For our main analysis, we use the following measures:

\begin{figure*}[!t]
    \centering
    \includegraphics[width = \textwidth]{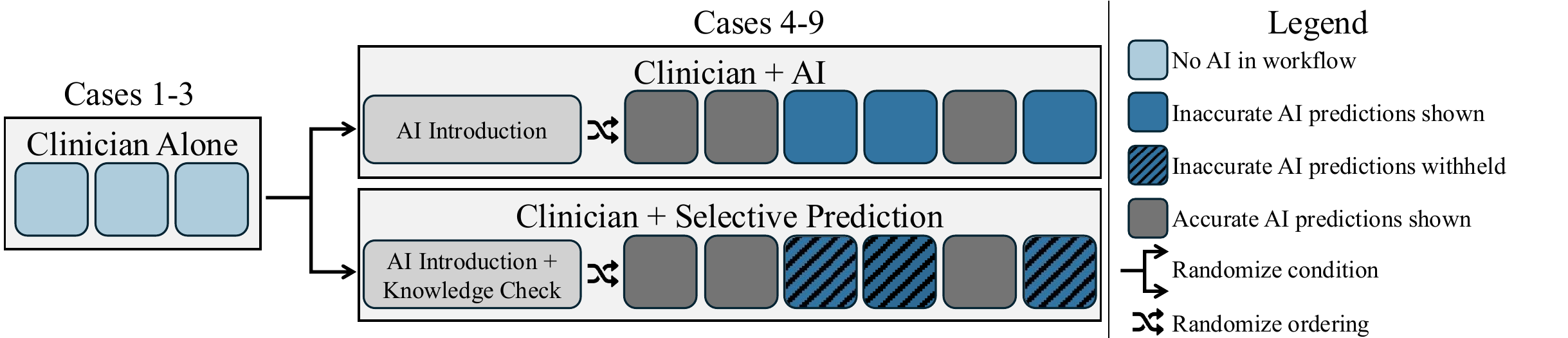}
   \caption{\textbf{Survey flow}. Clinicians first completed three patient cases without AI input, then were randomized to AI assistance with (\spspe) or without selective prediction (\ai). Both groups received an introduction to the AI model, and clinicians randomized to selective prediction completed a knowledge check to confirm understanding. The six AI-assisted cases included three from the \textit{inaccurate subset}, in which the selective prediction mechanism hid at least one of the predictions. Case order was randomized to mitigate ordering effects.}

    \label{fig:survey_flow}
\end{figure*}

\begin{itemize}
    \item \textbf{Average treatment accuracy}: We focus on clinicians' treatment accuracy in our main analysis, and include diagnostic accuracy in Appendix \ref{diagnostic_performance}. Each selected treatment was compared to the ground-truth diagnosis. A treatment decision was considered correct if it corresponded to the ground-truth diagnosis: antibiotics for pneumonia, intravenous diuretics for heart failure, and steroids for COPD. In this calculation, a decision to treat was positive, and negative otherwise. 
    \item  \textbf{Average false positive and false negative rates}: To understand how selective prediction changes clinician decision-making beyond accuracy, we analyze false positive and false negative rates for treatment decisions. 
\end{itemize}

\parnobf{Statistical Analysis.} All results reported in the main text are computed on the inaccurate subset. We use mixed-effects models to account for repeated decisions made by the same clinicians on the same patient cases~\citep{clayton1996generalized}. Since patients could have one, multiple, or none of the conditions, we treat each response for pneumonia, heart failure, and COPD as a separate data point. We fit a separate model for each outcome (accuracy, false negative rate, false positive rate), regressing the outcome on the experimental condition (\alone, \ai, \spspe) as a fixed effect and the patient, label, and participant ID as random effects. We report the p-values for each experimental condition with AI input (\ai, \spspe) relative to the baseline without AI in the workflow (\alone), using a significance level of 0.05.

\section{Experiments \& Results}

To understand the impact of selective prediction on clinical decisions related to treating ARF, we analyze the effects of inaccurate AI (\ai) and selective prediction (\spspe) on clinicians' treatment accuracy and types of errors clinicians make in each setting. 

\parnobf{Participants.} 259 clinicians completed the survey across nine US states (Appendix Table \ref{tab:demographics}). 125 were randomized to AI without selective prediction (\ai) and 134 were randomized to AI with selective prediction (\spspe). Both groups had a median 5 years of practice, and 95\% reported practicing hospital medicine, indicating that we successfully surveyed those who typically care for patients with ARF. 

\subsection{How does inaccurate AI with and without selective prediction affect treatment accuracy?}

To understand the effects of inaccurate AI and selective prediction, we measure clinicians’ treatment decisions in three settings: (1) \alone: No AI predictions shown;  (2) \ai: Inaccurate AI predictions shown; (3) \spspe: Inaccurate AI predictions withheld. We report numbers for all settings in Appendix Table \ref{tab:full_treatment_numbers}. 
\begin{figure*}[!t]
    \centering
    \begin{minipage}[c]{0.36\textwidth}
        \centering
        \includegraphics[width=\linewidth]{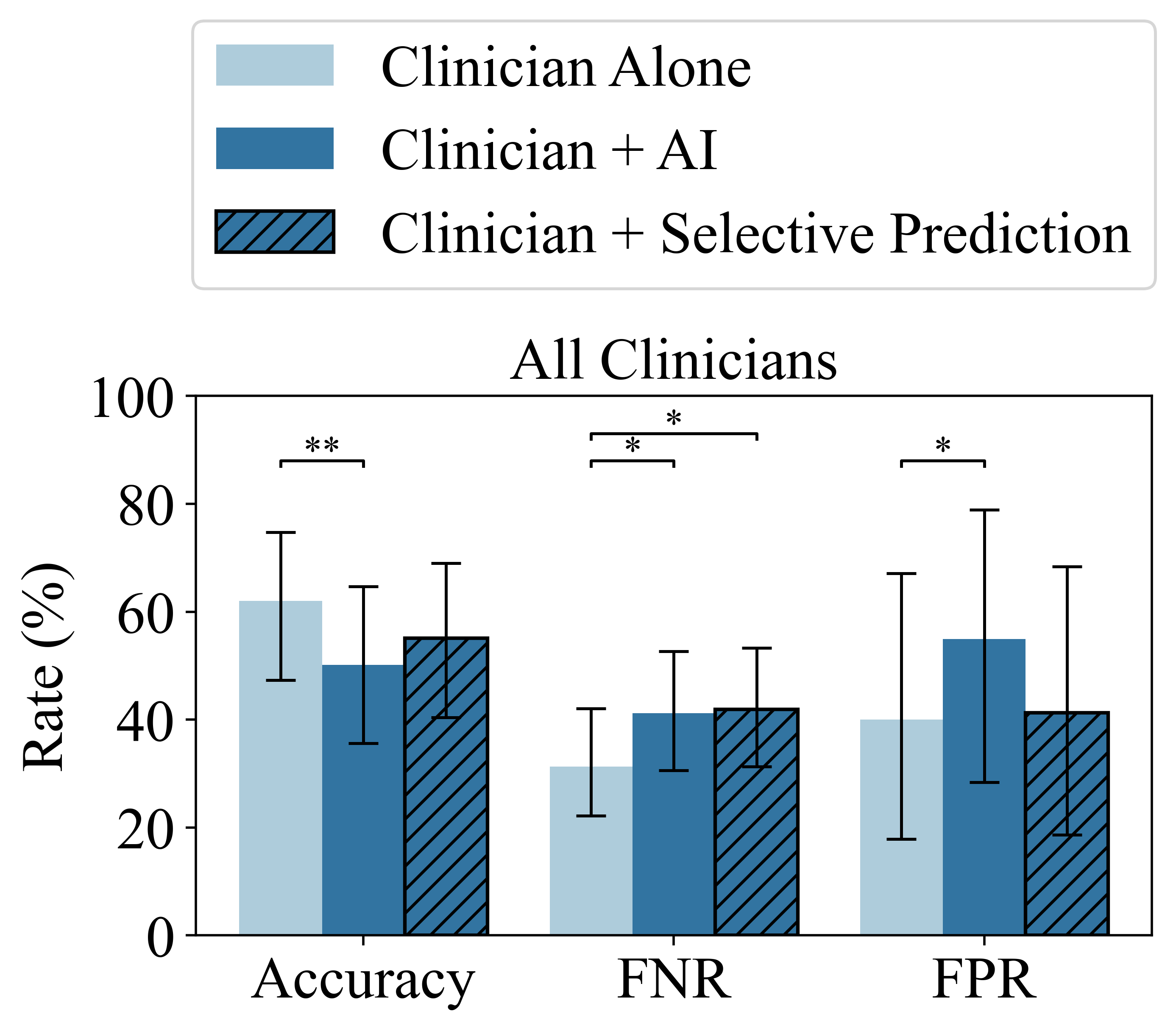}\\[2pt]
        {\small (a) Performance across all participants.}
    \end{minipage}\hfill
    \begin{minipage}[c]{0.62\textwidth}
        \centering
        \includegraphics[width=\linewidth]{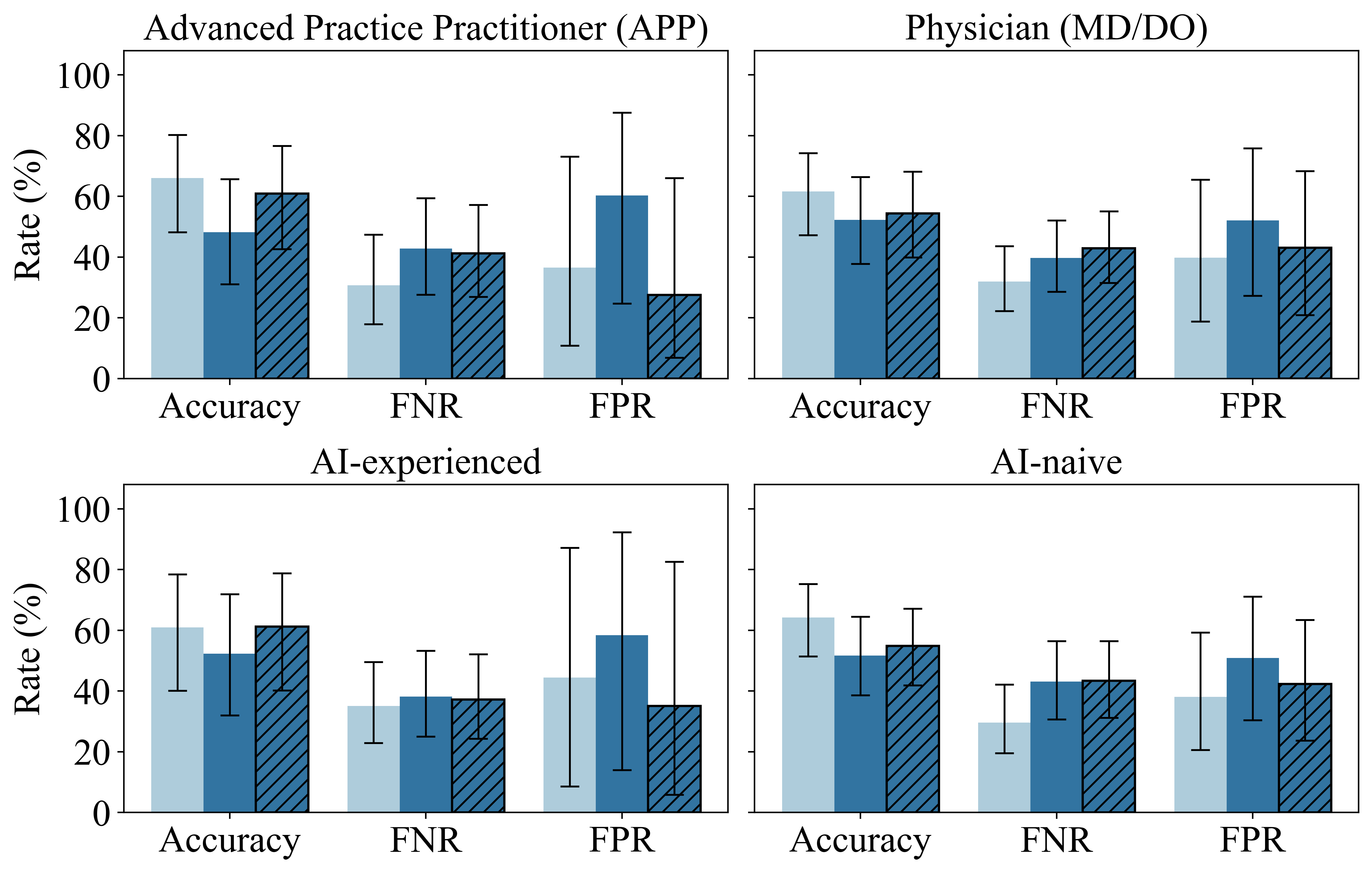}\\[2pt]
        {\small (b) Performance stratified by training and AI experience. }
    \end{minipage}

    \caption{\textbf{Treatment accuracy, false negative rate (FNR), and false positive rate (FPR) across experimental conditions with 95\% confidence intervals.} (a)~Overall results across all clinicians: In \ai, treatment accuracy drops significantly compared to \alone, and both false positive and false negative rates increase significantly. \spspe{} partially restores accuracy and brings the false positive rate back to baseline; however, the false negative rate remains significantly elevated ($*\,p<0.05$, $**\,p<0.01$). (b)~Exploratory analysis results stratified by clinician role (Advanced Practice Practitioners vs.\ Physicians, top row) and by prior AI use in clinical practice (bottom row). Top: trends held for both APPs and physicians, though effect magnitudes were larger for APPs. Bottom: the elevated false negative rate under \spspe{} held for AI-naive clinicians but not for AI-experienced clinicians.}
    \label{fig:overall_results}
\end{figure*}

\parnobf{Inaccurate AI hurts clinician accuracy.} In \alone, treatment accuracy was 62\% (95\% CI: [47-75]) (Figure \ref{fig:overall_results}a), matching our expectations given the difficulty of this task. In \ai, treatment accuracy dropped significantly relative to \alone{} (50\% [36-65] vs. 62\% [47-75]; p-value $<$ 0.001), suggesting overreliance on inaccurate predictions. 

This accuracy drop is driven by increases in both false positives and false negatives. Compared to \alone, false positives rose significantly from 40\% (18–67) to 55\% (28-79) (p-value $<$ 0.05) in \ai. False negatives similarly rose from 31\% (22–42) to 41\% (31–53) (p-value $<$ 0.05) (Figure~\ref{fig:overall_results}a). Trends were consistent in terms of diagnostic accuracy (Appendix \ref{diagnostic_performance}).

\parnobf{Selective prediction partly recovers clinician accuracy degraded by inaccurate AI.} Compared to \ai{}, treatment accuracy was higher under selective prediction (55\% [40–69] vs. 50\% [36–65]; Figure~\ref{fig:overall_results}a). Accuracy remains slightly lower than \alone{} (55\% [40–69] vs. 62\% [47–75]), though this difference is not statistically significant, suggesting that selective prediction partly, but not fully, mitigates the harm of inaccurate AI.

This recovery is driven by a partial recovery in clinicians' error patterns. Under \spspe, clinicians' false positive rate returned to baseline levels observed in \alone{} (41\% [19–68] vs. 40\% [18–67]) (Figure \ref{fig:overall_results}a). However, their false negative rate remained significantly elevated compared to \alone{} (42\% [31–53] vs. 31\% [22–42]; p-value $<$ 0.05), explaining why accuracy only \textit{partly} recovers. This indicates that clinicians are more likely to undertreat patients who need intervention when the AI abstains. Trends were consistent in terms of diagnostic decision errors, although point estimates for false positives were slightly lower under \spspe~compared to \alone~(Appendix \ref{diagnostic_performance}). 

\section{Exploratory Analyses}

We conducted three exploratory analyses to better understand how selective prediction affects different participant groups and how it shapes participants’ perceptions of patient cases. These analyses are exploratory and intended to contextualize overall trends rather than support formal statistical comparisons between experimental conditions. Thus, we did not test for statistical significance because the study was not sufficiently powered for formal subgroup comparisons.

\subsection{Do the effects of selective prediction vary by clinicians' length of medical training?} We examined whether clinicians' length of medical training influenced their interaction with AI. We grouped participants into two categories: (1) \textit{Physicians:} those with a Doctor of Medicine (MD) or Doctor of Osteopathic Medicine (DO) degree (residents, fellows, or attendings), and (2) \textit{Advanced Practice Practitioners:} (APPs), physician assistants (PAs) and nurse practitioners (NPs), who undergo shorter medical training and do not hold an MD or DO. This distinction allowed us to examine how differences in training length relate to the effects of selective prediction. We split responses by clinician type (APP vs. physician), and fit mixed-effects models within each subgroup. 

\parnobf{AI has stronger effects on APPs than on physicians.} 66 APPs and 191 physicians completed the survey (Appendix Table \ref{tab:demographics}). Results for APPs and physicians were consistent with the overall trends observed in the main analysis, though the effects of AI were markedly stronger for APPs. In \ai, APP accuracy decreased by 18 percentage points (pp) relative to \alone, compared to a 10 pp decrease for physicians (Figure \ref{fig:overall_results}b). Under \spspe, APP accuracy recovered by 13 pp to within 5 pp of baseline, whereas physician accuracy recovered by only 2 pp. At the same time, under \spspe, false negative rates were elevated relative to \alone{} for both groups: APP false negative rates were 10 pp higher, and physician false negative rates were 11 pp higher.

\subsection{Do the effects of selective prediction vary by clinicians' prior AI exposure?} We also examined whether clinicians' prior interaction with AI tools in clinical care influenced their interaction with AI. At the end of the survey, participants were asked \textit{``Have you interacted with a clinical decision support tool before in your clinical practice?''} We grouped participants into two categories: (1) \textit{AI-experienced}: those who responded `Yes', and (2) \textit{AI-naive}: those who responded `No.' We again fit mixed-effects models within each subgroup.

\parnobf{AI has stronger effects on AI-naive clinicians.}  78 participants were AI-experienced, whereas 181 were AI-naive (Appendix Table \ref{tab:ai-experience}). The effects of selective prediction were stronger on AI-naive clinicians compared to AI-experienced clinicians. In \ai, accuracy decreased by 12 percentage points (pp) for AI-naive clinicians relative to \alone, compared to a 9 pp decrease for AI-experienced clinicians (Figure \ref{fig:overall_results}b). Under \spspe, accuracy recovered by only 3 pp for AI-naive clinicians, but recovered fully by 9 pp for AI-experienced clinicians. These trends were reflected in their false negative rates as well: under \spspe, AI-naive clinicians' false negative rates were elevated by 13 pp relative to \alone{}, whereas they were only elevated by 2 pp for AI-experienced clinicians. 

\begin{figure}[!t]
    \centering
    \includegraphics[width = \linewidth]{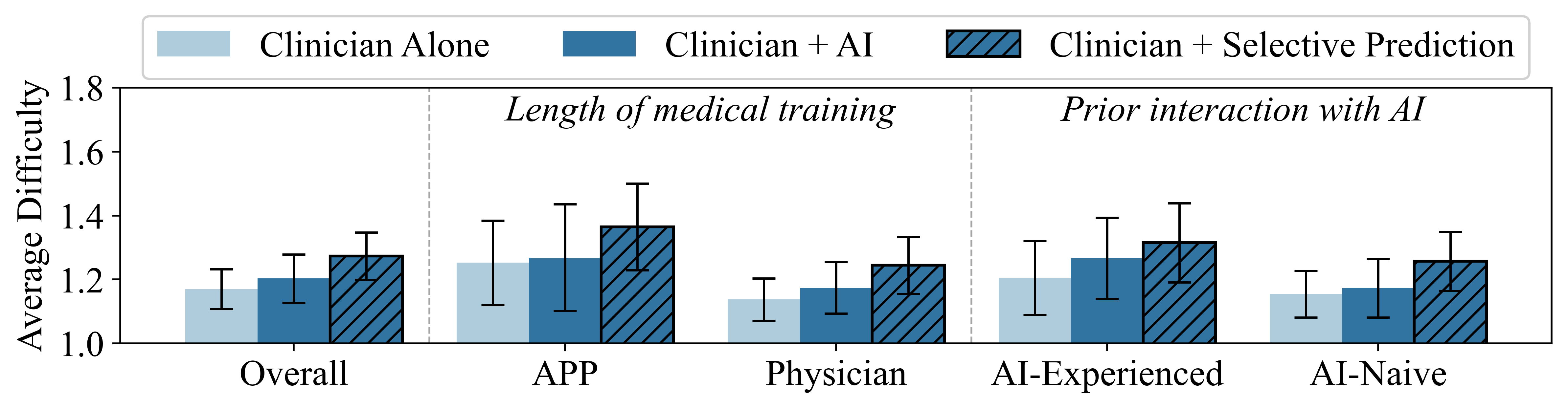}
    \caption{Average difficulty rating across settings on cases with inaccurate predictions. \ai{} increases the perceived difficulty, and \spspe{} increases this further.}
    \label{fig:difficulty}
\end{figure}

\subsection{Does selective prediction affect clinicians' perception of case difficulty?} We examined perceived case difficulty as a subjective indicator of how AI with or without selective prediction shapes clinicians’ experience of the task. We converted Likert-scale difficulty ratings into integers (1–4) and computed average difficulty ratings in each condition, with standard errors obtained via a nonparametric block bootstrap with resampling at the participant level~\citep{davison1997bootstrap}. 

\parnobf{Selective prediction increases perceived case difficulty.} Under selective prediction, clinicians perceive cases as more difficult than when no AI is present or when all AI predictions are shown. At baseline, average perceived case difficulty was 1.17 (95\% CI: [1.11-1.23]). \ai{} increased the perceived difficulty of patient cases to 1.20 (1.13-1.28) and \spspe{} increased this further to 1.27 (1.20-1.35) (Figure \ref{fig:difficulty}). Trends were consistent for both APPs and physicians, as well as AI-naive and AI-experienced clinicians. 
    
\section{Discussion \& Conclusion}

In this study, we tested a core assumption underlying the common evaluation of selective prediction: that when an AI system withholds a prediction and explicitly defers to the human, the human behaves as they would without any AI involvement in their workflow. Our results show that this assumption does not fully hold. While selective prediction partly mitigates the harmful effects of inaccurate AI on treatment accuracy, it also increases false negatives, leading to undertreatment relative to clinicians' baseline performance without AI. This is particularly pronounced amongst those who self-reported not to have previously used AI in their clinical practice. Importantly, these effects arise not from the accuracy of the selective prediction mechanism itself, but from clinicians' responses to AI abstention. When clinicians are informed that the model has withheld its prediction, abstention appears to influence decision-making, rather than functioning as a neutral absence of information.  

Prior work on selective prediction has largely focused on single-output binary decision settings, where AI input is either fully revealed or completely withheld. For example, Bondi \textit{et al.} showed in a conservation task that AI deferral can mitigate the harm of inaccurate AI~\citep{bondi2022role}. However, in their setting, participants either saw all AI predictions or none at all. In contrast, we study selective prediction in a multilabel setting, where the AI withholds a subset of predictions and the human is responsible for reasoning over \textit{incomplete} AI input. Our results suggest that the conclusions drawn from binary deferral settings do not directly transfer to multilabel decision-making. 

There are several potential reasons for the observed behavior in our study. First, clinicians may interpret AI deferral as the absence of a diagnosis rather than a neutral statement. We presented those randomized to selective prediction with a knowledge check, requiring them to confirm their understanding that the AI model was not expressing an opinion on the deferred conditions. However, we cannot be certain that participants fully understood selective prediction, and future work could focus on improving understanding. Second, clinicians may be subject to \textit{availability bias}, the tendency to weigh the likelihood of things by how easily they are recalled~\citep{gopal2021implicit}. Since the AI only presents a subset of predictions, clinicians might perceive those presented as more likely causes of patients' respiratory failure if they come to mind more easily. 

While general trends were consistent across participants grouped by length of medical training, AI appears to have stronger effects on those with less medical training (APPs) than those with more medical training (physicians). APPs typically work under physician supervision ~\citep{Smith_2023}, so they may rely more on AI due to lower confidence in their own judgments or tendency to agree with external diagnoses. In contrast, physicians are trained to critically question decisions~\citep{araujo2024critical} and may scrutinize AI outputs more closely. The effects of selective prediction also differed by prior AI experience: AI-naive clinicians were negatively influenced by selective prediction, but this was not observed in AI-experienced clinicians. One possible interpretation is that AI-experienced clinicians better understand AI's weaknesses, consistent with prior work showing that humans can learn to calibrate their trust in AI through experience~\citep{li2026learning}. However, AI experience was correlated with clinician demographics in our sample: AI-experienced clinicians were more likely to be male, older, and physicians rather than APPs (Appendix Table \ref{tab:ai_use_correlates}). Thus, the response of this group to selective prediction may reflect any of these correlated factors. Disentangling these effects would be an important direction for future research.

\parnobf{Limitations.} Our study has limitations. First, while we use a vignette-based setting test-bed, our online survey-based approach differs from real clinical environments. In practice, clinicians may feel greater pressure to make accurate diagnostic and treatment decisions, which might influence their reliance on AI. Further investigation in an integrated clinical setting is needed to determine if our findings generalize. However, such studies are not without risk, and our work provides an important first step in identifying potential deployment harms. Second, clinicians may have needed more training with the AI tool to understand selective prediction. Formal training could mitigate the observed error tradeoff in this study by ensuring participants recognize that deferred conditions should still be considered for diagnosis and treatment. 

We use clinical decisions as a test-bed to examine the effects of selective prediction in a multilabel setting, but our findings should raise questions beyond healthcare. AI systems that augment human decisions are often developed in a vacuum, optimized alone for retrospective performance. Yet, integrating AI into human workflows is just as important as optimizing AI accuracy alone.

\bibliographystyle{plainnat}
\bibliography{ref}

\appendix
\appendix

\clearpage
\section{Survey Setup} 
\label{survey}

\paragraph{Study Invitations.} We identified physicians at each hospital who sent an invitation email with the study information to hospitalist clinicians (those who treat patients with ARF) at their respective institutions.

\paragraph{Landing Page.} Once participants clicked on the study link, they were screened for eligibility. Specifically, they needed to confirm their role as a nurse practitioner, physician assistant, resident, fellow, attending physician, or any similar role. If they responded ``no," the study was terminated. Otherwise, they were redirected to an introduction page. On the introduction page, they were invited to participate in a research study aimed at understanding how clinicians might use clinical decision support tools in their diagnostic and treatment decisions. They were told they'd be presented with 9 cases based on real patient encounters that were de-identified, and would be asked about the likelihood of various diagnoses, what treatments they would provide, and how difficult they perceived the cases to be. They were also asked to give their best answer to each question and to treat each patient as if it were a real clinical encounter. They were informed that the study was anonymous, it would take about 20 minutes to complete, and that they could stop the study at any time and come back to the point where they left off.  So as to not bias their answers, they were also told that some details of the study's purpose would be withheld. They were also informed that they would receive a \$50 Amazon.com gift card upon completion of the study.

\begin{figure}[!htbp]
    \centering
    \includegraphics[width=0.9\columnwidth,height=0.8\textheight,keepaspectratio]{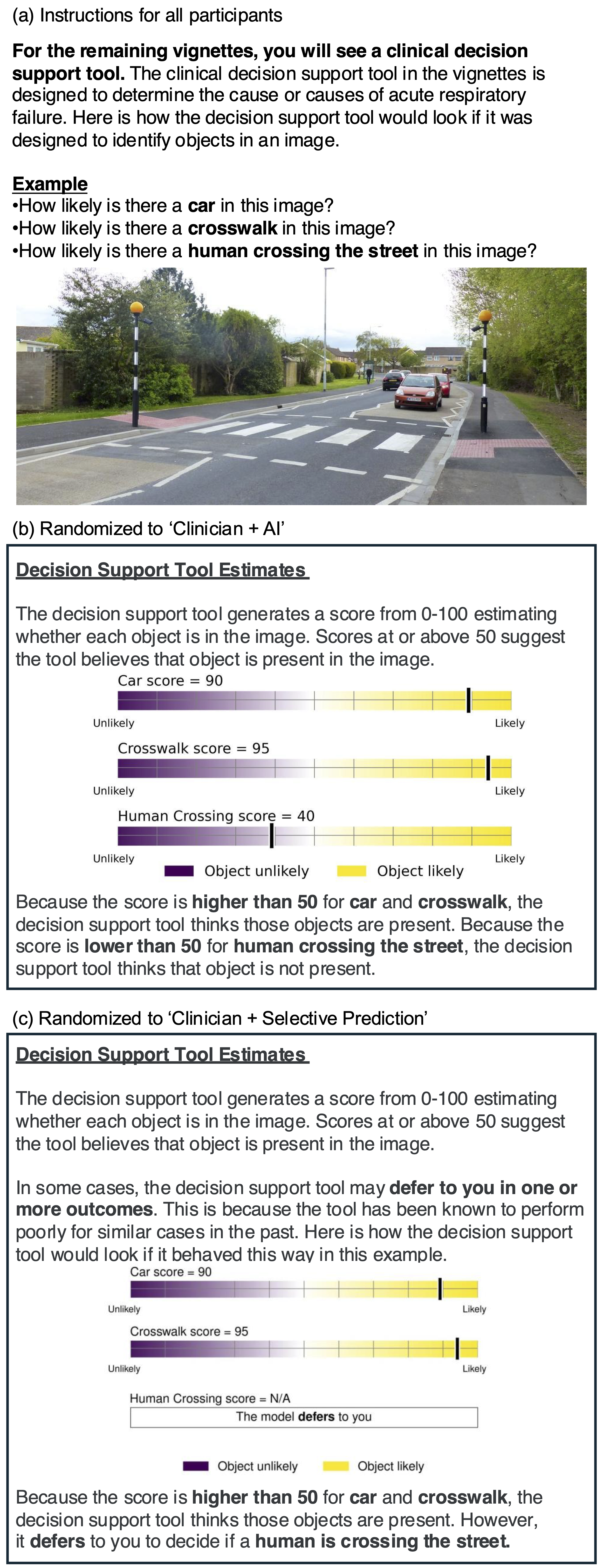}
    \caption{Instructions presented to participants about the AI model they would be presented with. (a) The model was described as a clinical decision support tool to all participants, and they were shown how it would behave if it were used to identify objects in consumer photographs. (b) If randomized to \ai, participants were shown how the model would present all predictions. (c) If randomized to \spspe, participants were shown how the model would behave if it deferred for a subset of the predictions.}
    \label{fig:instructions}
\end{figure}

\paragraph{Study Instructions.} Participants were told that they’d see a clinical decision support tool that was designed to determine the cause or causes of a patient's ARF. To not skew their perception of the tool before seeing it in the study, they were then shown an example of how the tool would work if it was designed to identify objects in consumer images (Figure \ref{fig:instructions}). 

\paragraph{IRB Approval.} This study was exempt from IRB review, as it was deemed human subjects research with adults that involve benign (non-harmful) behavioral interventions, information was collected through written responses, no physiological data was collected, subject identifiers were not collected, and the subjects agreed to participate in the intervention and information collection. However, we collected the patient data to generate clinical cases under a separate approved IRB protocol.

\begin{figure}[!htbp]
    \centering
    \includegraphics[width = \columnwidth]{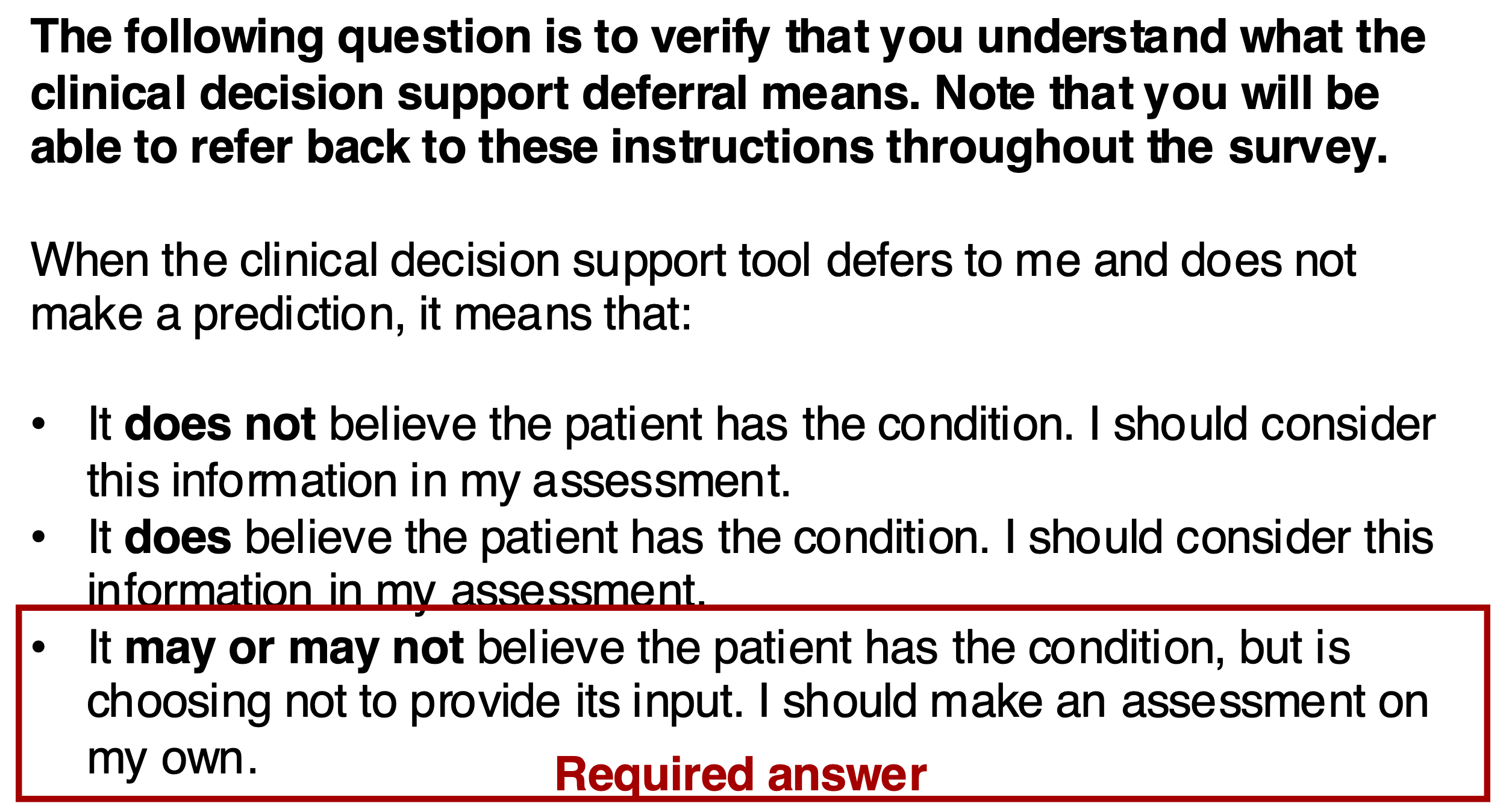}
    \caption{Knowledge check for those randomized to \spspe.}
    \label{fig:knowledge_check}
\end{figure}

\paragraph{Knowledge Check.} When randomized to \spspe, participants were required to answer a question to confirm their understanding of selective prediction (Figure \ref{fig:knowledge_check}). Participants were required to answer correctly, otherwise they couldn't continue the survey.

\paragraph{Survey Validation.} During survey pilot testing, we ensured the randomization worked as intended. However, after data collection was complete, we identified an issue with the selection of the first three clinical cases without AI input. While still randomly sampled, some participants encountered a patient case first without AI and then later with AI. To assess any potential impact, we reran our analyses excluding the second occurrence of the case and found no effect on our results. Therefore, we retained all case responses in our analyses for completeness.

\paragraph{Unusual Survey Responses.} While the survey was live, we observed 13 individuals completing it unusually quickly at a non-typical hour. Given this suspicious behavior, we contacted them to verify their medical training and eligibility but received no responses. To ensure our sample consisted of medical professionals caring for ARF patients, we excluded their responses. We still ensured we collected 250 total responses as our power calculations intended, separate from removing the 13 suspicious responses.

\section{Patient Cases}
\label{patient_cases}
\paragraph{Selecting Patient Cases.} We constructed 45 clinical patient cases based on real patients selected from a consecutive sample of 121 patients hospitalized with ARF at our institution between August and November of 2017. Patients used in the clinical cases were selected to ensure the sample was generally representative of patients hospitalized with ARF.

\paragraph{Ground Truth Diagnosis.} The ground truth diagnosis of each patient was determined by at least four physicians who typically care for ARF patients. They independently reviewed each patient's medical record to determine if pneumonia, heart failure, and/or COPD was the patient's underlying cause of the ARF on a scale of 1 to 4, with 1 being likely and 4 being unlikely. Scores were averaged and thresholded at 2.5 to indicate whether the patient had the disease, and patients could have any combination of the three (or none at all). In addition to determining the underlying cause of ARF (including pneumonia, heart failure, and/or COPD), each physician also gave a difficulty rating on a scale of 1-4 to describe the difficulty of diagnosing the specific patient case. Next, the 45 patients ultimately used in the study were selected from the larger group of 121 patients based on ensuring the sample was generally representative of patients hospitalized with ARF. This included ensuring the 45 patients selected had similar diagnostic difficulty across the three diagnoses studied (heart failure, pneumonia, and COPD) and similar diagnostic difficulty to the larger group of 121 patients. Finally, ensuring the 45 patients were selected to ensure disease prevalence for heart failure, pneumonia, and COPD was similar to the prevalence reported in prior national reports.

\paragraph{Case Development.} The clinical cases were developed iteratively by the study team members and piloted with 15 board certified internal medicine physicians not involved in the final study. First, a physician study team member reviewed the patient’s hospital admission history and physical exam and created an abbreviated version of each for the case. The patient’s chest X-ray image and laboratory data were also collected and added to the case.

A pilot study was conducted with the draft cases, AI model predictions, where clinicians provided qualitative feedback on study layout, case content, survey questions, and time required to complete each case. Improvements were made to the cases iteratively until clinicians consistently agreed that the case provided sufficient information to conduct the diagnostic assessments required during the study.

After incorporating all participant feedback from the initial pilot, a nurse clinical informaticist not involved in the research study reviewed each finalized case to ensure no identifiable patient
information was present. 

\section{Model details} 
\label{model_details}

To generate model predictions, we trained a machine learning model to predict pneumonia, heart failure, and COPD from the patients' chest X-rays and tabular clinical data. Two separate models were trained: one based on chest X-rays and the second on EHR data. To generate a final prediction for each patient, we then averaged the image- and EHR-based model predictions, so as to weight the chest radiographs and EHR data equally. 

We ensembled the predictions of two separate models based on the clinical data and chest X-rays as follows: 

\begin{enumerate}
    \item \textit{EHR model}: We trained a logistic regression and 2-layer neural network (1 hidden layer, size=100) with a sigmoid activation to estimate the probability of each diagnosis based on EHR data inputs, treating the model architecture as a hyperparameter. The best EHR model, either logistic regression or 2-layer neural network, was chosen based on validation AUROC performance. 
    \item \textit{Image model}: We trained a convolutional neural network (CNN) with a DenseNet-121 architecture to estimate the probability of each diagnosis based on the chest X-ray input. The model was first pretrained using X-rays from two publicly available datasets, CheXpert~\citep{irvin2019chexpert} and MIMIC-CXR~\citep{johnson2019mimic,johnson2024mimic,goldberger2000physiobank} to identify common radiographic findings annotated in radiology reports. Then the last layer of the model was fine-tuned to determine ARF diagnoses.
    
\end{enumerate}

Model predictions were shown as follows: model outputs were on a scale of 0-1. These outputs were then thresholded to yield a model decision. Predictions less than the threshold were deemed as the model predicting no disease, and predictions greater than the threshold as model predicting disease. Thresholds were set to the percent of positive patients for each disease using all the data except the 45 patients used to test the model. We measured the percent of positive patients for each disease in this dataset (pneumonia: 31\%, heart failure: 22\%, COPD: 8\%). For each diagnosis, we chose the threshold to be such that the percentage of patients labeled positive by the model was equal to the incidence rate of the diagnosis across this dataset. After thresholding, negative predictions were normalized to 0-50, and positive predictions were normalized to 51-100 so that the decision threshold shown to participants was the midpoint between 0 and 100. Scores were presented on a scale of 0-100, with 0 being unlikely and 100 being likely. 

\section{Subset Construction }
\label{app:subset_construction}

Specifically, the mechanism deferred in conditions where the model was highly inaccurate (i.e., where showing the prediction would likely harm clinical decisions) but did \textit{not} defer in highly accurate conditions (i.e., those conditions where showing the prediction was likely to help). The mechanism identifies most, but not all, harmful cases: it sometimes defers even when the model prediction is correct. Using an oracle separates the mechanism's accuracy from its effect. We identified these conditions as follows: 
\begin{itemize}
    \item \textbf{Inaccurate Subset: Model gets it (mostly) wrong}. Of the 135 predictions, we identified 28 predictions (14 pneumonia, 8 heart failure, and 6 COPD) where the model achieved an average accuracy of 10\%. These predictions corresponded to 23 patient cases (for some patients, the model was inaccurate for multiple conditions). For these conditions, the model \textit{could} abstain from providing the prediction under selective prediction, if the participant was randomized to such behavior. 
    \item \textbf{Accurate Subset: Model gets it (mostly) right}. The model gets 107 of 135 predictions mostly right, with an average accuracy of 91\%. These model predictions are always shown to participants, no matter the randomization. Forty-one of the predictions came from the same 23 patients as the inaccurate subset (totaling 69 predictions for this set of 23 patients), while 66 predictions came from the other 22 patients. 
    
\end{itemize}
While we could have used only patient cases where the AI defers on at least one condition, we recognize this would be unrealistic. In practice, clinicians are more likely to encounter a mix of cases: some where the AI defers on a subset of conditions (cases with both inaccurate and accurate predictions) and others where it defers on none (cases with only accurate predictions). Thus, we included both types of cases in the survey.

\section{Randomization} 
\label{randomization}

We developed a block randomization procedure to assign specific clinical cases to study participants. The block randomization procedure was used to randomize study participants to see AI model predictions with or without selective prediction. At the same time, it also ensured that all 45 patient cases would be evenly assigned across the first three baseline cases, and ensured that a subject would only see a specific clinical case once during the survey.

The randomization was performed in blocks of 90 to achieve the two randomizations (AI with and without selective prediction) This ensured that all 45 patient cases would be evenly assigned across the first three baseline cases, and to ensure that a subject would only see a clinical case once during the survey. Within the blocks of 90, 45 subjects were randomly assigned to each randomization. There were six cases where the AI model was displayed. Three cases contained both inaccurate and accurate predictions and were sampled from 23 patients. The other three cases contained only accurate predictions coming from 22 patients. Therefore, the three cases that included inaccurate predictions were sampled from the corresponding 23 patients such that each patient was seen 11 or 12 times within the block. The three cases that only include accurate predictions were sampled from the corresponding 22 patients such that each patient was seen 12 or 13 times within the block. These six patient cases were then displayed in a random order. 

\begin{enumerate}
    \item \textbf{Cases 4-9: AI Input}. The next six cases included AI input, each with three AI predictions generated for each condition. In a realistic setting, AI models will not always defer to clinicians. Thus, three patient cases were chosen from the set of 23 patients containing inaccurate and accurate predictions. The other three cases were chosen from the set of 22 patients with only accurate predictions. We maintained a 50/50 ratio of patient cases with and without inaccurate predictions to ensure sufficient power to test our hypotheses. The order of cases with and without selective prediction was randomized to avoid ordering effects. Participants were randomized into two groups: seeing \textit{all} AI predictions or seeing AI with selective prediction. Thus, while participants saw both inaccurate and accurate predictions, whether or not selective prediction was applied depended on the randomization. 

\end{enumerate}

\section{Power Calculations} 

The primary goal of the study was to understand the impact of inaccurate AI and AI with selective prediction on clinician accuracy. Answering these questions required making specific diagnostic accuracy comparisons across 3 experimental settings (\alone), (\ai), (\spspe). The power calculation focused on ensuring adequate power to detect differences in accuracy across these experimental settings. In the \ai{} setting, participants were shown a lower accuracy prediction for the diagnosis. When calculating diagnostic accuracy in the \ai{} setting, only inaccurate predictions were used from the three cases in which inaccurate predictions were shown. For this reason, the power calculations were designed to ensure adequate power in detecting the effects of \ai{} and \spspe{} with respect to inaccurate predictions. We aimed to recruit 250 participants via simulation-based power calculations, which ensured adequate power to detect a 10\% decrease in diagnostic accuracy with highly inaccurate AI, and then a recovery to baseline accuracy with selective prediction.

\section{End-of-Survey Questions}
\label{end_of_survey_demographics}

We collected optional demographic information (Section \ref{app:demographics_questions}) and asked questions about participants' experiences and attitudes towards AI (Section \ref{app:experience_questions}) after they completed all cases in the study. We report responses to these results in Table \ref{tab:demographics} and \ref{tab:ai-experience}.

\clearpage
\subsection{Demographics Questions}
\label{app:demographics_questions}

\begin{longtable}{@{}p{0.97\linewidth}@{}}
\toprule
\textbf{What is your age today (in years)?} \\[2pt]
\textit{Free-text entry (numeric).} \\
\midrule
\textbf{What is your race or ethnicity? Select all that apply.} \\[2pt]
$\square$~American Indian or Alaska Native \\
$\square$~Asian \\
$\square$~Black \\
$\square$~Hispanic or Latino \\
$\square$~Middle Eastern \\
$\square$~Native Hawaiian or Pacific Islander \\
$\square$~White \\
$\square$~Prefer not to say \\
$\square$~Other (please specify) \\
\midrule
\textbf{What is your gender?} \\[2pt]
$\square$~Male \\
$\square$~Female \\
$\square$~Non-binary / non-conforming \\
$\square$~Transgender \\
$\square$~Other (please specify) \\
$\square$~Prefer not to say \\
\midrule
\textbf{In what hospital setting do you primarily work? Select all that apply.} \\[2pt]
$\square$~University Hospital \\
$\square$~Community Hospital / Private Practice \\
$\square$~VA / Government \\
$\square$~Other (please specify) \\
\midrule
\textbf{What is your general practice area?} \\[2pt]
$\square$~Hospital Medicine \\
$\square$~Pulmonary / Critical Care Medicine \\
$\square$~Emergency Medicine \\
$\square$~Cardiology \\
$\square$~Infectious Disease \\
$\square$~Radiology \\
$\square$~Outpatient Primary Care \\
$\square$~Other (please specify) \\
\midrule
\textbf{What is your current role on the healthcare team?} \\[2pt]
$\square$~Nurse Practitioner (NP) \\
$\square$~Physician Assistant \\
$\square$~Resident \\
$\square$~Fellow \\
$\square$~Attending Physician \\
$\square$~Other \\
\midrule
\textbf{In which state do you currently reside?} \\[2pt]
\textit{Dropdown of 50 U.S. states, D.C., and Puerto Rico.} \\
\midrule
\textbf{In what year did you receive your medical degree?} \\[2pt]
\textit{Displayed if role is Resident or Fellow. Free-text (year).} \\
\midrule
\textbf{In what year did you complete residency?} \\[2pt]
\textit{Displayed if role is Attending Physician. Free-text (year).} \\
\midrule
\textbf{In what year did you complete fellowship (if applicable)?} \\[2pt]
\textit{Displayed if role is Attending Physician. Free-text (year).} \\
\midrule
\textbf{In what year did you complete your medical training?} \\[2pt]
\textit{Displayed if role is Nurse Practitioner or Physician Assistant. Free-text (year).} \\
\bottomrule
\end{longtable}

\subsection{Experience and Attitudes Toward Clinical Decision Support}
\label{app:experience_questions}
\begin{longtable}{@{}p{0.97\linewidth}@{}}
\toprule
\textbf{Have you interacted with a clinical decision support tool before in your clinical practice?} \\[2pt]
$\square$~Yes \\
$\square$~No \\
\midrule
\textbf{How often do you interact with a clinical decision support tool in your clinical practice?} \\[2pt]
\textit{Displayed if previous question answered Yes.} \\[2pt]
$\square$~Never \\
$\square$~Rarely (just specific patients with difficult cases) \\
$\square$~Sometimes (1--4 times for every 10 patients) \\
$\square$~Often (5--8 times for every 10 patients) \\
$\square$~Always (9--10 times for every 10 patients) \\
$\square$~Other (please specify) \\
\midrule
\textbf{How important are clinical decision support tools in your clinical practice?} \\[2pt]
\textit{Displayed if interaction with CDS = Yes.} \\[2pt]
$\square$~Very unimportant \\
$\square$~Somewhat unimportant \\
$\square$~Neutral \\
$\square$~Somewhat important \\
$\square$~Very important \\
$\square$~Other (please specify) \\
\midrule
\textbf{When you interact with a clinical decision support tool, how do you use the information provided by the tool?} \\[2pt]
\textit{Displayed if interaction with CDS = Yes.} \\[2pt]
$\square$~I always use the information in my clinical decision making \\
$\square$~I consider the information, but it is not always a part of my clinical decision making \\
$\square$~I completely ignore the information \\
$\square$~Other (please specify) \\
\midrule
\textbf{Have you ever recommended a clinical decision support tool to your colleagues?} \\[2pt]
\textit{Displayed if interaction with CDS = Yes.} \\[2pt]
$\square$~Yes \\
$\square$~No \\
\midrule
\textbf{To what extent do you agree with the following statement:} \textit{Using artificial intelligence to analyze medical images and diagnose disease could be beneficial to my patient care and management.} \\[2pt]
$\square$~Strongly agree \\
$\square$~Agree \\
$\square$~Neither agree nor disagree \\
$\square$~Disagree \\
$\square$~Strongly disagree \\
\bottomrule
\end{longtable}
\begin{table}[t]
\centering
\caption{Study participant demographics. Race/ethnicity and hospital setting allowed multiple selections, so percentages may sum to over 100\%.}
\small
\begin{tabular}{lrr}
\toprule
\textbf{Randomization} & \textbf{Clinician + AI} & \textbf{Clinician + Selective Prediction} \\
\midrule
Total Responses, n & 125 & 134 \\
Number Complete Responses, n (\%) & 125 (100.0) & 134 (100.0) \\
Included in regression analyses, n (\%) & 125 (100.0) & 134 (100.0) \\
Response Time (Minutes), median (IQR) & 19 (14-28) & 20 (15-32) \\
Age, median (IQR) & 35 (32-39) & 35 (31-41) \\
Years of Practice, median (IQR) & 5 (2-10) & 5 (2-9) \\
\midrule
\textbf{Gender, n (\%)} & & \\
\quad Female & 66 (52.8) & 72 (53.7) \\
\quad Male & 54 (43.2) & 61 (45.5) \\
\quad Prefer not to say & 5 (4.0) & 1 (0.7) \\
\midrule
\textbf{Race/Ethnicity, n (\%)} & & \\
\quad White & 72 (57.6) & 71 (53.0) \\
\quad Asian & 35 (28.0) & 38 (28.4) \\
\quad Prefer not to say & 12 (9.6) & 10 (7.5) \\
\quad Middle Eastern & 5 (4.0) & 8 (6.0) \\
\quad Hispanic or Latinx & 3 (2.4) & 5 (3.7) \\
\quad Black & 0 (0.0) & 5 (3.7) \\
\quad Other & 1 (0.8) & 1 (0.7) \\
\midrule
\textbf{State of Practice, n (\%)} & & \\
\quad Massachusetts & 44 (35.2) & 54 (40.3) \\
\quad Michigan & 27 (21.6) & 21 (15.7) \\
\quad Colorado & 13 (10.4) & 13 (9.7) \\
\quad Texas & 13 (10.4) & 13 (9.7) \\
\quad Wisconsin & 6 (4.8) & 12 (9.0) \\
\quad Ohio & 13 (10.4) & 11 (8.2) \\
\quad California & 8 (6.4) & 8 (6.0) \\
\quad Missouri & 0 (0.0) & 1 (0.7) \\
\quad North Carolina & 1 (0.8) & 0 (0.0) \\
\midrule
\textbf{General Practice Area, n (\%)} & & \\
\quad Hospital Medicine & 119 (95.2) & 128 (95.5) \\
\quad Other & 6 (4.8) & 5 (3.7) \\
\quad Cardiology & 0 (0.0) & 1 (0.7) \\
\midrule
\textbf{Role on Healthcare Team, n (\%)} & & \\
\quad Attending Physician & 87 (69.6) & 98 (73.1) \\
\quad Physician Assistant & 27 (21.6) & 25 (18.7) \\
\quad Nurse Practitioner (NP) & 8 (6.4) & 6 (4.5) \\
\quad Resident/Fellow & 2 (1.6) & 4 (3.0) \\
\quad Other & 1 (0.8) & 0 (0.0) \\
\midrule
\textbf{Hospital Setting, n (\%)} & & \\
\quad University Hospital/Academic & 115 (92.0) & 117 (87.3) \\
\quad Community Hospital/Private Practice & 11 (8.8) & 19 (14.2) \\
\quad VA/Government & 7 (5.6) & 4 (3.0) \\
\bottomrule
\end{tabular}
\label{tab:demographics}
\end{table}

\begin{table}[t]
\centering
\caption{Study participant prior experience with and attitudes toward AI.}

\small
\begin{tabular}{lrr}
\toprule
\textbf{Randomization} & \textbf{Clinician + AI} & \textbf{Clinician + Selective Prediction} \\
\midrule
\textbf{Used AI in clinical practice, n (\%)} & & \\
\quad No & 84 (67.2) & 97 (72.4) \\
\quad Yes & 41 (32.8) & 37 (27.6) \\
\midrule
\textbf{Interaction frequency, n (\%) -- AI users only} & & \\
\quad Never & 1 (2.4) & 2 (5.4) \\
\quad Rarely (just specific patients with difficult cases) & 11 (26.8) & 16 (43.2) \\
\quad Sometimes (1-4 times for every 10 patients) & 25 (61.0) & 18 (48.6) \\
\quad Often (5-8 times for every 10 patients)  & 2 (4.9) & 0 (0.0) \\
\quad Always (9-10 times for every 10 patients) & 1 (2.4) & 1 (2.7) \\
\quad Other & 1 (2.4) & 0 (0.0) \\
\midrule
\textbf{Importance in decisions, n (\%) -- AI users only} & & \\
\quad Very unimportant & 2 (4.9) & 3 (8.1) \\
\quad Somewhat unimportant & 4 (9.8) & 9 (24.3) \\
\quad Neutral & 10 (24.4) & 13 (35.1) \\
\quad Somewhat important & 18 (43.9) & 10 (27.0) \\
\quad Very important & 5 (12.2) & 2 (5.4) \\
\quad Other & 2 (4.9) & 0 (0.0) \\
\midrule
\textbf{Recommends AI to colleagues, n (\%) -- AI users only} & & \\
\quad Yes & 21 (51.2) & 17 (45.9) \\
\quad No & 20 (48.8) & 20 (54.1) \\
\midrule
\textbf{AI could benefit patient care, n (\%)} & & \\
\quad Strongly disagree & 1 (0.8) & 2 (1.5) \\
\quad Disagree & 10 (8.0) & 7 (5.2) \\
\quad Neither agree nor disagree & 29 (23.2) & 31 (23.1) \\
\quad Agree & 66 (52.8) & 72 (53.7) \\
\quad Strongly agree & 19 (15.2) & 22 (16.4) \\
\bottomrule
\end{tabular}
\label{tab:ai-experience}
\end{table}

\newpage 
\clearpage
\section{Performance in the inaccurate subset}
\label{supp_spe}

Here we report the numerical values of all Figures \ref{fig:overall_results} in the main text for improved understanding of our results.

\subsection{Numeric Values for Treatment Performance}
\begin{table}[!h]
\centering
\caption{Treatment accuracy, FPR, and FNR by subgroup. Significance markers shown for the Overall column only. Key: $p<0.05$: *; $p<0.01$: **.}

\resizebox{\textwidth}{!}{%
\begin{tabular}{lccccc}
\toprule
 & Overall & APP & Physician & AI-experienced & AI-naive \\
\midrule
\multicolumn{6}{l}{\textbf{Accuracy}} \\
Clinician Alone & 0.62 (0.47-0.75) & 0.66 (0.48-0.80) & 0.62 (0.47-0.74) & 0.61 (0.40-0.78) & 0.64 (0.51-0.75) \\
Clinician + AI & 0.50 (0.36-0.65)** & 0.48 (0.31-0.66) & 0.52 (0.38-0.66) & 0.52 (0.32-0.72) & 0.52 (0.39-0.64) \\
Clinician + Selective Prediction & 0.55 (0.40-0.69) & 0.61 (0.43-0.77) & 0.54 (0.40-0.68) & 0.61 (0.40-0.79) & 0.55 (0.42-0.67) \\
\midrule
\multicolumn{6}{l}{\textbf{False Positive Rate}} \\
Clinician Alone & 0.40 (0.18-0.67) & 0.36 (0.11-0.73) & 0.40 (0.19-0.65) & 0.44 (0.09-0.87) & 0.38 (0.21-0.59) \\
Clinician + AI & 0.55 (0.28-0.79)* & 0.60 (0.25-0.88) & 0.52 (0.27-0.76) & 0.58 (0.14-0.92) & 0.51 (0.30-0.71) \\
Clinician + Selective Prediction & 0.41 (0.19-0.68) & 0.28 (0.07-0.66) & 0.43 (0.21-0.68) & 0.35 (0.06-0.83) & 0.42 (0.24-0.63) \\
\midrule
\multicolumn{6}{l}{\textbf{False Negative Rate}} \\
Clinician Alone & 0.31 (0.22-0.42) & 0.31 (0.18-0.47) & 0.32 (0.22-0.44) & 0.35 (0.23-0.50) & 0.30 (0.20-0.42) \\
Clinician + AI & 0.41 (0.31-0.53)* & 0.43 (0.28-0.59) & 0.40 (0.29-0.52) & 0.38 (0.25-0.53) & 0.43 (0.31-0.56) \\
Clinician + Selective Prediction & 0.42 (0.31-0.53)* & 0.41 (0.27-0.57) & 0.43 (0.32-0.55) & 0.37 (0.24-0.52) & 0.43 (0.31-0.56) \\
\bottomrule
\end{tabular}%
}

\label{tab:full_treatment_numbers}
\end{table}

\subsection{Correlation between AI experience and participant demographics}

We measured the correlation between participants' self-reported AI use in clinical care and their demographics include age, years practicing medicine, gender, and role in clinical care.

\begin{table}[!h]
\centering
\small
\caption{Exploratory analysis reporting associations between clinician demographics and self-reported prior AI use in clinical practice. Continuous variables report median [IQR] and Mann--Whitney U test. Categorical variables report $n$ (\%) and Fisher's exact or chi-square test as appropriate. Multi-select variables (race/ethnicity, hospital setting) test each option independently.}
\label{tab:ai_use_correlates}
\begin{tabular}{@{}lllllr@{}}
\toprule
\textbf{Variable} & \textbf{Test} & \textbf{AI-experienced)} & \textbf{AI-naive} & \textbf{Statistic} & \textbf{$p$-value} \\
\midrule
\textbf{Age (years)} & Mann-Whitney U & 38 [32, 42] & 34 [31, 39] & U=7773 & \textbf{0.008} \\
\textbf{Years practicing} & Mann-Whitney U & 7 [2, 11] & 4 [2, 9] & U=7675 & 0.147 \\
\textbf{Gender:} Female & Chi-square (df=2) & 31 (39.7\%) & 107 (59.1\%) & chi2=9.69 & \textbf{0.008} \\
\quad Male &  & 46 (59.0\%) & 69 (38.1\%) &  &  \\
\quad Prefer not to say &  & 1 (1.3\%) & 5 (2.8\%) &  &  \\
\textbf{Role:} APP & Fisher's exact & 13 (16.7\%) & 57 (31.5\%) & -- & \textbf{0.015} \\
\quad Clinician &  & 65 (83.3\%) & 124 (68.5\%) &  &  \\

\bottomrule
\end{tabular}
\end{table}

\newpage

\subsection{Diagnostic Performance}
\label{diagnostic_performance}
We also measured diagnostic performance for participants, including diagnostic accuracy, false positive rates, and false negative rates. For each individual response, we compared the clinicians' diagnostic assessment to the ground truth diagnostic label. The diagnosis was considered correct if it agreed with the ground truth label. Responses of 50 or higher were considered positive for the diagnosis. Trends were consistent with our main analyses.

\begin{table}[!h]
\centering
\caption{Diagnosis accuracy, FPR, and FNR overall and by subgroup for the inaccurate subset. Significance markers shown for the Overall column only. Key: $p<0.05$: *; $p<0.01$: **.}
\resizebox{\textwidth}{!}{%
\begin{tabular}{lccccc}
\toprule
 & Overall & APP & Physician & AI-experienced & AI-naive \\
\midrule
\multicolumn{6}{l}{\textbf{Accuracy}} \\
Clinician Alone & 0.63 (0.49-0.76) & 0.68 (0.43-0.86) & 0.64 (0.52-0.74) & 0.64 (0.47-0.78) & 0.65 (0.51-0.76) \\
Clinician + AI & 0.53 (0.39-0.67)** & 0.45 (0.22-0.70) & 0.58 (0.45-0.69) & 0.55 (0.37-0.71) & 0.55 (0.41-0.68) \\
Clinician + Selective Prediction & 0.62 (0.47-0.74) & 0.59 (0.34-0.81) & 0.62 (0.50-0.73) & 0.64 (0.47-0.79) & 0.62 (0.48-0.74) \\
\midrule
\multicolumn{6}{l}{\textbf{False Positive Rate}} \\
Clinician Alone & 0.28 (0.14-0.47) & 0.26 (0.07-0.63) & 0.26 (0.14-0.45) & 0.25 (0.11-0.48) & 0.27 (0.13-0.49) \\
Clinician + AI & 0.44 (0.25-0.65)** & 0.56 (0.21-0.86) & 0.39 (0.22-0.60) & 0.44 (0.23-0.67) & 0.42 (0.22-0.64) \\
Clinician + Selective Prediction & 0.21 (0.10-0.39) & 0.21 (0.05-0.59) & 0.21 (0.11-0.38) & 0.18 (0.07-0.38) & 0.24 (0.11-0.44) \\
\midrule
\multicolumn{6}{l}{\textbf{False Negative Rate}} \\
Clinician Alone & 0.39 (0.30-0.48) & 0.33 (0.17-0.53) & 0.41 (0.32-0.50) & 0.40 (0.28-0.53) & 0.39 (0.29-0.49) \\
Clinician + AI & 0.44 (0.35-0.54) & 0.49 (0.29-0.69) & 0.43 (0.34-0.53) & 0.39 (0.27-0.53) & 0.46 (0.36-0.57) \\
Clinician + Selective Prediction & 0.46 (0.37-0.56) & 0.44 (0.26-0.64) & 0.48 (0.39-0.57) & 0.45 (0.32-0.59) & 0.47 (0.36-0.57) \\
\bottomrule
\end{tabular}%
}
\end{table}

\section{Difficulty}
\label{supp_diff}
We report numerical values for Figure \ref{fig:difficulty} in the main text.

\begin{table}[!h]
\centering
\caption{Average difficulty rating and corresponding 95\% confidence intervals on the inaccurate subset. Participant difficulty rating increased when shown all AI predictions (\ai) and increased further when shown AI with selective prediction (\spspe)}
\resizebox{\textwidth}{!}{%
\begin{tabular}{lccccc}
\toprule
 & Overall & APP & Physician & AI-Experienced & AI-Naive \\
\midrule
Clinician Alone & 1.17 (1.11-1.23) & 1.25 (1.12-1.38) & 1.14 (1.07-1.20) & 1.20 (1.09-1.32) & 1.15 (1.08-1.23) \\
Clinician + AI & 1.20 (1.13-1.28) & 1.27 (1.10-1.44) & 1.17 (1.09-1.25) & 1.27 (1.14-1.39) & 1.17 (1.08-1.26) \\
Clinician + Selective Prediction & 1.27 (1.20-1.35) & 1.36 (1.23-1.50) & 1.24 (1.16-1.33) & 1.32 (1.19-1.44) & 1.26 (1.16-1.35) \\
\bottomrule
\end{tabular}%
}
\end{table}

\newpage

\section{Performance in the accurate subset}
\label{supp_spi}

Here, we report participant performance on responses to the accurate subset of conditions. We make two comparisons. First, we measure clinician performance at baseline on the accurate subset (\alone). Then, we compare to two groups: clinician performance on the accurate subset when randomized to AI without selective prediction (\ai) and clinician performance on the accurate subset when randomized to AI with selective prediction (\spspe) (though note that they still see all predictions for the accurate subset when the AI defers on the inaccurate subset). 

\noindent\textbf{Clinician treatment accuracy improves with accurate AI predictions.} At baseline (\textit{Clinician Alone}), clinician treatment accuracy on the accurate subset without AI input is 76\% (63--86) (Table~\ref{tab:acc_subset_treatment}). Clinician accuracy increases significantly when presented accurate predictions both without (\textit{Clinician + AI} = 81\% [69--89]) and with (\textit{Clinician + Selective Prediction} = 83\% [71--90]) selective prediction. Improvements in accuracy are also reflected in lower false positive and false negative rates in both settings (\textit{Clinician + AI}, \textit{Clinician + Selective Prediction}), though false negative differences from baseline (\textit{Clinician Alone}) are not statistically significant. However, clinician performance remains below that of the AI model (91\%). Trends are consistent across clinician role (APP vs. Physician), AI experience (AI-experienced vs. AI-naive), as well as with diagnostic accuracy measurements (Table \ref{tab:acc_subset_diagnosis}). 

\begin{table}[!htbp]
    \centering
    \caption{Treatment accuracy for the accurate subset in the different settings with 95\% confidence intervals. Key: p-value $<$ 0.05: $*$; p-value $<$ 0.01: $**$.}

    \resizebox{\textwidth}{!}{%
\begin{tabular}{lccccc}
\toprule
 & Overall & APP & Physician & AI-experienced & AI-naive \\
\midrule
\multicolumn{6}{l}{\textbf{Accuracy}} \\
Clinician Alone & 0.76 (0.63-0.86) & 0.77 (0.60-0.88) & 0.76 (0.63-0.86) & 0.77 (0.65-0.86) & 0.76 (0.61-0.86) \\
Clinician + AI & 0.81 (0.69-0.89)** & 0.82 (0.67-0.91) & 0.81 (0.69-0.89) & 0.82 (0.72-0.90) & 0.80 (0.67-0.89) \\
Clinician + Selective Prediction & 0.83 (0.71-0.90)** & 0.84 (0.69-0.92) & 0.83 (0.72-0.90) & 0.85 (0.75-0.91) & 0.82 (0.70-0.90) \\
\midrule
\multicolumn{6}{l}{\textbf{False Positive Rate}} \\
Clinician Alone & 0.23 (0.11-0.41) & 0.24 (0.11-0.46) & 0.23 (0.12-0.41) & 0.23 (0.13-0.38) & 0.23 (0.11-0.44) \\
Clinician + AI & 0.17 (0.08-0.33)** & 0.17 (0.07-0.35) & 0.17 (0.08-0.33) & 0.19 (0.10-0.32) & 0.17 (0.07-0.35) \\
Clinician + Selective Prediction & 0.16 (0.07-0.31)** & 0.15 (0.06-0.32) & 0.16 (0.08-0.31) & 0.16 (0.08-0.27) & 0.16 (0.07-0.33) \\
\midrule
\multicolumn{6}{l}{\textbf{False Negative Rate}} \\
Clinician Alone & 0.15 (0.05-0.34) & 0.11 (0.04-0.28) & 0.16 (0.07-0.34) & 0.12 (0.05-0.25) & 0.15 (0.05-0.38) \\
Clinician + AI & 0.13 (0.04-0.30) & 0.10 (0.03-0.27) & 0.14 (0.05-0.30) & 0.08 (0.03-0.17) & 0.14 (0.05-0.37) \\
Clinician + Selective Prediction & 0.11 (0.04-0.27) & 0.10 (0.03-0.27) & 0.12 (0.05-0.26) & 0.07 (0.03-0.17) & 0.11 (0.04-0.31) \\
\bottomrule
\end{tabular}%
}
    \label{tab:acc_subset_treatment}
\end{table}

\begin{table}[!h]
\centering
\caption{Diagnosis accuracy, FPR, and FNR by subgroup for the accurate subset. Significance markers shown for the Overall column only. Key: $p<0.05$: *; $p<0.01$: **.}
\resizebox{\textwidth}{!}{%
\begin{tabular}{lccccc}
\toprule
 & Overall & APP & Physician & AI-experienced & AI-naive \\
\midrule
\multicolumn{6}{l}{\textbf{Accuracy}} \\
Clinician Alone & 0.80 (0.69-0.88) & 0.76 (0.63-0.85) & 0.82 (0.70-0.90) & 0.80 (0.68-0.89) & 0.80 (0.68-0.88) \\
Clinician + AI & 0.86 (0.76-0.92)** & 0.82 (0.71-0.90) & 0.87 (0.77-0.93) & 0.89 (0.80-0.94) & 0.84 (0.73-0.91) \\
Clinician + Selective Prediction & 0.86 (0.77-0.92)** & 0.84 (0.74-0.91) & 0.87 (0.78-0.93) & 0.89 (0.81-0.94) & 0.85 (0.75-0.91) \\
\midrule
\multicolumn{6}{l}{\textbf{False Positive Rate}} \\
Clinician Alone & 0.18 (0.09-0.31) & 0.24 (0.13-0.39) & 0.15 (0.08-0.28) & 0.18 (0.10-0.31) & 0.18 (0.09-0.32) \\
Clinician + AI & 0.12 (0.06-0.23)** & 0.16 (0.09-0.29) & 0.11 (0.05-0.21) & 0.10 (0.05-0.19) & 0.14 (0.07-0.27) \\
Clinician + Selective Prediction & 0.12 (0.06-0.22)** & 0.14 (0.07-0.26) & 0.11 (0.05-0.20) & 0.09 (0.05-0.18) & 0.12 (0.06-0.24) \\
\midrule
\multicolumn{6}{l}{\textbf{False Negative Rate}} \\
Clinician Alone & 0.15 (0.05-0.38) & 0.14 (0.05-0.34) & 0.15 (0.05-0.38) & 0.15 (0.06-0.33) & 0.15 (0.05-0.39) \\
Clinician + AI & 0.11 (0.03-0.29) & 0.10 (0.03-0.27) & 0.10 (0.03-0.29) & 0.08 (0.03-0.19) & 0.12 (0.04-0.33) \\
Clinician + Selective Prediction & 0.10 (0.03-0.28) & 0.09 (0.03-0.26) & 0.10 (0.03-0.28) & 0.09 (0.03-0.22) & 0.11 (0.04-0.31) \\
\bottomrule
\end{tabular}%
}
    \label{tab:acc_subset_diagnosis}

\end{table}

\newpage
\clearpage

\end{document}